# Universal relaxation times for electron and nucleon gases

M. Pelc[*]

Institute of Physics, Maria Curie – Sklodowska University, Lublin, Poland

J. Marciak – Kozlowska

Institute of Electron Technology, Warsaw, Poland

M. Kozłowski

Institute of Experimental Physics, Warsaw University, Warsaw, Poland

---

[*] corresponding author: magdaanna@o2.pl

Abstract

In this paper we calculate the universal relaxation times for electron and nucleon fermionic gases. We argue that the universal relaxation time tau(i) is equal tau(i)=h/m square v(i) where v(i)=alpha(i)c and alpha(1)=0.15 for nucleon gas and alpha(2)=1/137 for electron gas, c=light velocity. With the universal relaxation time we formulate the thermal Proca equation for fermionic gases.

Introduction

The differential equations of thermal energy transfer should be hyperbolic so as to exclude action at distance; yet the equations of irreversible thermodynamics – those of Navier – Stokes and Fourier are parabolic.

In the book [1] the new hyperbolic non – Fourier equation for heat transport was formulated and solved.

The excitation of matter on the quark nuclear and atomic level leads to transfer of energy. The response of the chunk of matter (nucleus, atom) is governed by the relaxation time.

In this paper we develop the general, universal definition of the relaxation time, which depends on coupling constants for electromagnetic or strong interaction.

It occurs that the general formula for the relaxation time can be written as

$$\tau_i = \frac{\hbar}{m_i(\alpha_i c)^2} \qquad (1)$$

where $m_i$ is the heat carrier mass, $\alpha_i = \left(i = e, 1/137, i = N, m_\pi/m_n\right)$ is coupling constant for electromagnetic and strong interaction, $c$ is the vacuum light speed. As the $c$ is the maximal velocity all relaxation time fulfils the inequality

$$\tau > \tau_i$$

Consequently $\tau_i$ is the minimal universal relaxation time.

1. Quantum heat transport equation

Dynamical processes are commonly investigated using laser pump-probe experiments with a pump pulse exciting the system of interest and a second probe pulse tracking is temporal evolution. As the time resolution attainable in such experiments depends on the temporal definition of the laser pulse, pulse compression to the attosecond domain is a recent promising development.

After the standards of time and space were defined the laws of classical physics relating such parameters as distance, time, velocity, temperature are assumed to be independent of accuracy with which these parameters can be measured. It should be noted that this assumption does not enter explicitly into the formulation of classical physics. It implies that together with the assumption of existence of an object and really independently of any measurements (in classical physics) it was tacitly assumed that *there was a possibility of an unlimited increase in accuracy of measurements*. Bearing in mind the "atomicity" of time i.e. considering the smallest time period, the Planck time, the above statement is obviously not true. Attosecond laser pulses we are at the limit of laser time resolution.

With attosecond laser pulses belong to a new Nano – World where size becomes comparable to atomic dimensions, where transport phenomena follow different laws from that in the macro world. This first stage of miniaturization, from $10^{-3}$ m to $10^{-6}$ m is over and the new one, from $10^{-6}$ m to $10^{-9}$ m just beginning. The Nano – World is a quantum world with all the predicable and non-predicable (yet) features.

In this paragraph, we develop and solve the quantum relativistic heat transport equation for nanoscale transport phenomena where external forces exist [2]. In paragraph 2 we developethe new hyperbolic heat transport equation which generalizes the Fourier heat transport equation for the rapid thermal processes. The hyperbolic heat transport equation (HHT) for the fermionic system has be written in the form (3):

$$\frac{1}{\left(\frac{1}{3}\upsilon_F^2\right)}\frac{\partial^2 T}{\partial t^2} + \frac{1}{\tau\left(\frac{1}{3}\upsilon_F^2\right)}\frac{\partial T}{\partial t} = \nabla^2 T \quad , \tag{3}$$

where $T$ denotes the temperature, $\tau$ the relaxation time for the thermal disturbance of the fermionic system, and $\upsilon_F$ is the Fermi velocity.

In what follows we present the formulation of the HHT, considering the details of the two fermionic systems: electron gas in metals and the nucleon gas [1].

For the electron gas in metals, the Fermi energy has the form

$$E_F^e = (3\pi)^2 \frac{n^{2/3}\hbar^2}{2m_e}, \tag{4}$$

where $n$ denotes the density and $m_e$ electron mass. Considering that

$$n^{-1/3} \sim a_B \sim \frac{\hbar^2}{me^2}, \tag{5}$$

and $a_B$ = Bohr radius, one obtains

$$E_F^e \sim \frac{n^{2/3}\hbar^2}{2m_e} \sim \frac{\hbar^2}{ma^2} \sim \alpha^2 m_e c^2, \tag{6}$$

where $c$ = light velocity and $\alpha = 1/137$ is the fine-structure constant for electromagnetic interaction. For the Fermi momentum $p_F$ we have

$$p_F^e \sim \frac{\hbar}{a_B} \sim \alpha m_e c, \tag{7}$$

and, for Fermi velocity $\upsilon_F$,

$$\upsilon_F^e \sim \frac{p_F}{m_e} \sim \alpha c. \tag{8}$$

Formula (8) gives the theoretical background for the result presented in paragraph 2 Considering formula (8), equation HHT can be written as

$$\frac{1}{c^2}\frac{\partial^2 T}{\partial t^2} + \frac{1}{c^2\tau}\frac{\partial T}{\partial t} = \frac{\alpha^2}{3}\nabla^2 T. \tag{9}$$

As seen from (9), the HHT equation is a relativistic equation, since it takes into account the finite velocity of light.

For the nucleon gas, Fermi energy equals

$$E_F^N = \frac{(9\pi)^{2/3} \hbar^2}{8mr_0^2}, \tag{10}$$

where $m$ denotes the nucleon mass and $r_0$, which describes the range of strong interaction, is given by

$$r_0 = \frac{\hbar}{m_\pi c}, \tag{11}$$

wherein $m_\pi$ is the pion mass. From formula (11), one obtains for the nucleon Fermi energy

$$E_F^N \sim \left(\frac{m_\pi}{m}\right)^2 mc^2. \tag{12}$$

In analogy to the Eq. (6), formula (12) can be written as

$$E_F^N \sim \alpha_s^2 mc^2, \tag{13}$$

where $\alpha_s = \frac{m_\pi}{m} \cong 0.15$ is the fine-structure constant for strong interactions. Analogously, we obtain the nucleon Fermi momentum

$$p_F^e \sim \frac{\hbar}{r_0} \sim \alpha_s mc \tag{14}$$

and the nucleon Fermi velocity

$$\upsilon_F^N \sim \frac{pF}{m} \sim \alpha_s c, \tag{15}$$

and HHT for nucleon gas can be written as

$$\frac{1}{c^2}\frac{\partial^2 T}{\partial t^2} + \frac{1}{c^2 \tau}\frac{\partial T}{\partial t} = \frac{\alpha_s^2}{3}\nabla^2 T. \tag{16}$$

In the following, the procedure for the discretization of temperature $T(\vec{r},t)$ in hot fermion gas will be developed. First of all, we introduce the reduced de Broglie wavelength

$$\lambda_B^e = \frac{\hbar}{m_e \upsilon_h^e}, \qquad \upsilon_h^e = \frac{1}{\sqrt{3}}\alpha c,$$
$$\lambda_B^N = \frac{\hbar}{m \upsilon_h^N}, \qquad \upsilon_h^N = \frac{1}{\sqrt{3}}\alpha_s c, \tag{17}$$

and the mean free paths $\lambda^e$ and $\lambda^N$

$$\lambda^e = \upsilon_h^e \tau^e, \qquad \lambda^N = \upsilon_h^N \tau^N. \tag{18}$$

In view of formulas (17) and (18), we obtain the HHC for electron and nucleon gases

$$\frac{\lambda_B^e}{\upsilon_h^e} \frac{\partial^2 T}{\partial t^2} + \frac{\lambda_B^e}{\lambda^e} \frac{\partial T}{\partial t} = \frac{\hbar}{m_e} \nabla^2 T^e, \tag{19}$$

$$\frac{\lambda_B^N}{\upsilon_h^N} \frac{\partial^2 T}{\partial t^2} + \frac{\lambda_B^N}{\lambda^N} \frac{\partial T}{\partial t} = \frac{\hbar}{m} \nabla^2 T^N. \tag{20}$$

Equations (19) and (20) are the hyperbolic partial differential equations which are the master equations for heat propagation in Fermi electron and nucleon gases. In the following, we will study the quantum limit of heat transport in the fermionic systems. We define the quantum heat transport limit as follows:

$$\lambda^e = \lambdabar_B^e, \qquad \lambda^N = \lambdabar_B^N. \tag{21}$$

In that case, Eqs. (19) and (20) have the form

$$\tau^e \frac{\partial^2 T^e}{\partial t^2} + \frac{\partial T^e}{\partial t} = \frac{\hbar}{m_e} \nabla^2 T^e, \tag{22}$$

$$\tau^N \frac{\partial^2 T^N}{\partial t^2} + \frac{\partial T^N}{\partial t} = \frac{\hbar}{m} \nabla^2 T^N, \tag{23}$$

where

$$\tau^e = \frac{\hbar}{m_e (\upsilon_h^e)^2}, \qquad \tau^N = \frac{\hbar}{m (\upsilon_h^N)^2}. \tag{24}$$

Equations (22) and (23) define the master equation for quantum heat transport (QHT). Having the relaxation times $\tau^e$ and $\tau^N$, one can define the "pulsations" $\omega_h^e$ and $\omega_h^N$

$$\omega_h^e = (\tau^e)^{-1}, \qquad \omega_h^N = (\tau^N)^{-1}, \tag{25}$$

or

$$\omega_h^e = \frac{m_e (\upsilon_h^e)^2}{\hbar}, \qquad \omega_h^N = \frac{m (\upsilon_h^N)^2}{\hbar},$$

i.e.,

$$\omega_h^e \hbar = m_e (v_h^e)^2 = \frac{m_e \alpha^2}{3} c^2,$$
$$\omega_h^N \hbar = m (v_h^N)^2 = \frac{m \alpha_s^2}{3} c^2. \quad (26)$$

The formulas (26) define the Planck-Einstein relation for heat quanta $E_h^e$ and $E_h^N$

$$E_h^e = \omega_h^e \hbar = m_e (v_h^e)^2,$$
$$E_h^N = \omega_h^N \hbar = m_N (v_h^N)^2. \quad (27)$$

In Table 1 the hierarchy of the relaxation times $\tau_i$ is presented.

Table 1. Hierarchical relaxation times

| Structure | $\alpha_i$ | $\tau_i$, $i = N, e$ | Numerical values |
|---|---|---|---|
| Atomic nucleus | 0.15 | $\dfrac{\hbar}{m_N (\alpha_2 c)^2}$ | $10^{-23}$ s |
| Atom | 1/137 | $\dfrac{\hbar}{m_e (\alpha_1 c)^2}$ | $10^{-17}$ s |

The heat quantum with energy $E_h = \hbar\omega$ can be named the *heaton*, in complete analogy to the *phonon*, *magnon*, *roton*, etc. For $\tau^e, \tau^N \to 0$, Eqs. (22) and (26) are the Fourier equations with quantum diffusion coefficients $D^e$ and $D^N$

$$\frac{\partial T^e}{\partial t} = D^e \nabla^2 T^e, \qquad D^e = \frac{\hbar}{m_e}, \quad (28)$$

$$\frac{\partial T^N}{\partial t} = D^N \nabla^2 T^N, \qquad D^N = \frac{\hbar}{m}. \quad (29)$$

For finite $\tau^e$ and $\tau^N$, for $\Delta t < \tau^e$, $\Delta t < \tau^N$, Eqs. (22) and (23) can be written as

$$\frac{1}{(v_h^e)^2} \frac{\partial^2 T^e}{\partial t^2} = \nabla^2 T^e, \quad (30)$$

$$\frac{1}{(v_h^N)^2} \frac{\partial^2 T^N}{\partial t^2} = \nabla^2 T^N. \quad (31)$$

Equations (30) and (31) are the wave equations for quantum heat transport (QHT). For $\Delta t > \tau$, one obtains the Fourier equations (28) and (29).

2. Proca thermal equation

It is quite interesting that the Proca type equation can be obtained for thermal phenomena. In the following starting with the hyperbolic heat diffusion equation the Proca equation for thermal processes will be developed and solved [2].

In paper [2] the relativistic hyperbolic transport equation was developed:

$$\frac{1}{\upsilon^2}\frac{\partial^2 T}{\partial t^2} + \frac{m_0 \gamma}{\hbar}\frac{\partial T}{\partial t} = \nabla^2 T. \tag{32}$$

In equation (32) $\upsilon$ is the velocity of heat waves, $m_0$ is the mass of heat carrier and $\gamma$ – the Lorentz factor, $\gamma = \left(1 - \frac{\upsilon^2}{c^2}\right)^{-1/2}$. As was shown in paper [2] the heat energy (*heaton temperature*) $T_h$ can be defined as follows:

$$T_h = m_0 \gamma \upsilon^2. \tag{33}$$

Considering that $\upsilon$, the thermal wave velocity equals [2]

$$\upsilon = \alpha c, \tag{34}$$

where $\alpha$ is the coupling constant for the interactions which generate the *thermal wave* ($\alpha = 1/137$ and $\alpha = 0.15$ for electromagnetic and strong forces respectively), the *heaton temperature* is equal to

$$T_h = \frac{m_0 \alpha^2 c^2}{\sqrt{1-\alpha^2}}. \tag{35}$$

Based on equation (35) one concludes that the *heaton temperature* is a linear function of the mass $m_0$ of the heat carrier. It is interesting to observe that the proportionality of $T_h$ and the heat carrier mass $m_0$ was observed for the first time in ultrahigh energy heavy ion reactions measured at CERN [3]. In paper [3] it was shown that the temperature of pions, kaons and protons produced in Pb+Pb, S+S reactions are proportional to the mass of particles. Recently, at Rutherford Appleton Laboratory (RAL), the VULCAN LASER was used to produce the elementary particles: electrons and pions [4].

When the external force is present $F(x,t)$ the forced damped heat transport is obtained [2] (in one dimensional case):

$$\frac{1}{v^2}\frac{\partial^2 T}{\partial t^2} + \frac{m_0\gamma}{\hbar}\frac{\partial T}{\partial t} + \frac{2Vm_0\gamma}{\hbar^2}T - \frac{\partial^2 T}{\partial x^2} = F(x,t). \tag{36}$$

The hyperbolic relativistic quantum heat transport equation, (36), describes the forced motion of heat carriers which undergo scattering ($\frac{m_0\gamma}{\hbar}\frac{\partial T}{\partial t}$ term) and are influenced by the potential term ($\frac{2Vm_o\gamma}{\hbar^2}T$).

Equation (36) is the Proca thermal equation and can be written as [2]:

$$\left(\bar{\Box}^2 + \frac{2Vm_0\gamma}{\hbar^2}\right)T + \frac{m_0\gamma}{\hbar}\frac{\partial T}{\partial t} = F(x,t),$$

$$\bar{\Box}^2 = \frac{1}{v^2}\frac{\partial^2}{\partial t^2} - \frac{\partial^2}{\partial x^2}. \tag{37}$$

We seek the solution of equation (37) in the form

$$T(x,t) = e^{-t/2\tau} u(x,t), \tag{38}$$

where $\tau_i = \frac{\hbar}{mv^2}$ is the relaxation time. After substituting equation (38) in equation (37) we obtain a new equation

$$\left(\bar{\Box}^2 + q\right)u(x,t) = e^{t/2\tau} F(x,t) \tag{39}$$

and

$$q = \frac{2Vm}{\hbar^2} - \left(\frac{mv}{2\hbar}\right)^2, \tag{40}$$

$$m = m_0\gamma. \tag{41}$$

In free space i.e. when $F(x,t) \to 0$ equation (39) reduces to

$$\left(\bar{\Box}^2 + q\right)u(x,t) = 0, \tag{42}$$

which is essentially the free Proca type equation.

The Proca equation describes the interaction of the laser pulse with the matter. As was shown in book [1] the quantization of the temperature field leads to the *heatons* – quanta of thermal energy with a mass $m_h = \dfrac{\hbar}{\tau v_h^2}$ [1], where $\tau$ is the relaxation time and $v_h$ is the finite velocity for heat propagation. For $v_h \to \infty$, i.e. for $c \to \infty$, $m_0 \to 0$, it can be concluded that in non-relativistic approximation ($c$ = infinite) the Proca equation is the diffusion equation for massless photons and heatons.

3. Solution of the Proca thermal equation

For the initial *Cauchy* condition:

$$u(x,0) = f(x), \qquad u_t(x,0) = g(x) \tag{43}$$

the solution of the Proca equation has the form (for $q > 0$) [2]

$$\begin{aligned}
u(x,t) &= \frac{f(x-vt) + f(x+vt)}{2} \\
&+ \frac{1}{2v} \int_{x-vt}^{x+vt} g(\varsigma) J_0\left[\sqrt{q(v^2 t^2 - (x-\varsigma)^2)}\right] d\varsigma \\
&- \frac{\sqrt{q} v t}{2} \int_{x-vt}^{x+vt} f(\varsigma) \frac{J_1\left[\sqrt{q(v^2 t^2 - (x-\varsigma)^2)}\right]}{\sqrt{v^2 t^2 - (x-\varsigma)^2}} d\varsigma \\
&+ \frac{1}{2v} \int_0^t \int_{x-v(t-t')}^{x+v(t-t')} G(\varsigma,t') J_0\left[\sqrt{q(v^2(t-t')^2 - (x-\varsigma)^2)}\right] dt' d\varsigma.
\end{aligned} \tag{44}$$

where $G = e^{t/2\tau} F(x,t)$.

When $q < 0$ solution of Proca equation has the form:

$$\begin{aligned}
u(x,t) &= \frac{f(x-vt) + f(x+vt)}{2} \\
&+ \frac{1}{2v} \int_{x-vt}^{x+vt} g(\varsigma) I_0\left[\sqrt{-q(v^2 t^2 - (x-\varsigma)^2)}\right] d\varsigma \\
&- \frac{\sqrt{-q} v t}{2} \int_{x-vt}^{x+vt} f(\varsigma) \frac{I_1\left[\sqrt{-q(v^2 t^2 - (x-\varsigma)^2)}\right]}{\sqrt{v^2 t^2 - (x-\varsigma)^2}} d\varsigma \\
&+ \frac{1}{2v} \int_0^t \int_{x-v(t-t')}^{x+v(t-t')} G(\varsigma,t') I_0\left[\sqrt{-q(v^2(t-t')^2 - (x-\varsigma)^2)}\right] dt' d\varsigma.
\end{aligned} \tag{45}$$

When $q = 0$ equation (39) is the forced thermal equation

$$\frac{1}{v^2} \frac{\partial^2 u}{\partial t^2} - \frac{\partial^2 u}{\partial x^2} = G(x,t). \tag{46}$$

In this paper we developed the relativistic thermal transport equation for an attosecond laser pulse interaction with matter. It is shown that the equation obtained is the Proca equation, well known in relativistic electrodynamics for massive photons. As the *heatons* are massive particles the analogy is well founded.

Conclusions

In the paper [5] S. Hod has derived the universal bound on the relaxation time of a perturbated system, $\tau > \dfrac{\hbar}{\pi T}$ . Considering that $T$ – temperature for nonrelativistic particles is $T \sim mv^2$, where $v$ is the average velocity, Hod's formula is $\tau > \dfrac{\hbar}{mv^2}$, which is the same as our formula (24). It seems that in our case the formula for the minimal relaxation time is more specific and better suited for comparison to experiment.